\documentstyle[12pt,psfig,aasms4]{article}

\lefthead{GOODWIN, PEARCE \& THOMAS}
\righthead{SIMULATING SUPERNOVAE REMNANTS}

\begin{document}

\title{SIMULATING SUPERNOVAE REMNANTS IN GAS CLOUDS}

\author{SIMON\,P.\,GOODWIN\altaffilmark{1}}
\affil{Astronomy Centre, CPES, University of Sussex, Falmer, Brighton, BN1 9QJ}

\author{F.\,R.\,PEARCE\altaffilmark{2}}
\affil{Physics Department, Science Labs, University of Durham, Durham,
DH1 3LE}

\author{PETER\,A.\,THOMAS\altaffilmark{3}}
\affil{Astronomy Centre, CPES, University of Sussex, Falmer, Brighton, BN1 9QJ}

\altaffiltext{1}{spg@astr.cpes.susx.ac.uk}
\altaffiltext{2}{F.R.Pearce@durham.ac.uk}
\altaffiltext{3}{P.A.Thomas@sussex.ac.uk}
\begin{abstract}

The Hydra $N$-body hydrodynamics code has been modified to model, from 
the end of the Sedov phase, the effects of supernovae on the surrounding 
medium.  The motivation is to investigate the feedback of energy into 
the interstellar/intergalactic medium.  We compare our results for 
supernova remnants (SNRs) in a uniform medium to previous detailed work on 
the late evolution of SNRs.  The code is found to reproduce the bulk 
characteristics of SNRs well.  Results on the effects of a single central 
SNR on Plummer clouds are presented.  The feedback of kinetic energy and 
the percentage mass loss can be parameterised in terms of the cloud mass 
and characteristic radius in a simple way.

The kinetic energy fraction returned to the ISM from a SNR is $<3$ 
per cent.  The removal of gas from the cold, dense phase and the 
addition of energy due to the lowering of the potential energy of a cloud 
is at least as significant, if not much more so, than the kinetic energy 
leaving a cloud.

\end{abstract}


\section{INTRODUCTION}

The mechanism by which energy and metals are fed back into the inter-stellar 
medium (ISM) by the supernovae of massive stars is an interesting and 
complex problem with applications to many areas of astrophysics.  In this 
paper we present an adaptation of the Hydra $N$-body hydrodynamics 
code (Couchman, Thomas \& Pearce 1995) which is capable of modelling the 
effects of multiple interacting supernovae over a large range of scales.

Hydrodynamical simulations of galaxy evolution and cosmological structure 
formation have recently been a topic of particular interest (eg. Katz 
1992; Katz, Hernquist \& Weinberg 1992, 1999; Navarro \& White 1993; 
Steinmetz \& Muller 1995; Frenk et al. 1996; Pearce et al. 1999).  Feedback 
from supernovae appears to be required to reheat gas and prevent the `cooling 
catastrophe' (White \& Rees 1978; White \& Frenk 1991), caused because
cooling is very efficient in small halos where the cooling time is
less than the dynamical time. As the particle mass in a
simulation increases, more and more of these halos are resolved and a
greater fraction of the available gas cools, in contradiction with the
observation that only a small fraction of baryonic material is in the
form of stars or cold gas.  Presently the efficiency at which energy is 
returned to the ISM and beyond to the intergalactic medium (IGM) is a 
free parameter with little more than handwaving arguments to support the 
values adopted (eg. Katz 1992; Navarro \& White 1993; Mihos \& Hernquist 
1994; Gerritsen \& Icke 1997).  It is to  
be expected that such efficiency parameters depend upon the density and 
metallicity of the surrounding gas at the very least.  One of the 
main aims of this project is to better constrain the values 
of these feedback parameters (the amount of energy and material returned to 
the ISM as well as the spatial extent and form of the feedback) for a wide 
range of stellar systems.

An understanding of the early evolution of star clusters requires knowledge 
of how the residual gas, left within a cluster after star formation has 
occurred, is expelled (eg. Tenorio-Tagle et al. 1986; Goodwin 1997a, b).  Such 
an understanding would also help provide constraints on 
the initial conditions and formation models of star clusters (Goodwin 
1997b).  Within a larger context, information on the effects of 
multiple supernovae is required to construct detailed stimulated star 
formation models.

Much detailed work (both hydrodynamic simulations and analytic calculations) 
has been carried out upon the dynamics of individual supernovae expanding 
into a uniform ISM (see Lozinskaya 1992 and references therein), however, 
little of this work has concentrated upon the late phases of that 
expansion (exceptions including Cioffi, McKee \& Bertschinger 1988; Slavin 
\& Cox 1992; Thornton et al. 1998).  Even less has been published on the 
effects of supernovae in gas clouds; analytic calculations were made of 
the effects of multiple supernovae on gas clouds by Dopita \& Smith (1986) 
and Morgan \& Lake (1989), but due to the complexity of the problem 
the treatment was understandably simplistic.  More recently Petruk 
(1999a, b) has published simulations of the very early evolution of a 
SNR in a density gradient.

In this paper we describe the extensions we have made to the Hydra code 
to model this problem and present a number of convergence tests that we 
have performed.  We also compare the results from this code to previous 
analytic and simulation work on individual supernovae.  In section 2 we 
describe the testing of the code and show that it 
replicates the behaviour of a supernova remnant expanding into a uniform 
medium as studied by other authors.  In section 3 we study the effects of 
a central supernova in Plummer clouds of various masses and characteristic 
radii.  In section 4 conclusions are drawn and the future applications of 
the code detailed.

\section{TESTING}

In this section we present the code we have used and test both its 
self-consistency and it's ability to reproduce known results for the 
evolution of SNRs in a uniform density ISM.  Firstly we outline 
the modifications to the Hydra code.  Then we show that the evolution of 
the energy content of an SNR and its expansion law is in good agreement 
with other simulations.   We show that these laws hold for different 
ambient densities and that the properties of an SNR scale with its 
initial radius.  

We do this to show that we are confident of the code's ability to model 
the evolution of a SNR when applied to the problems of SNR evolution in a 
cloud in Section 3 (and in future papers).

\subsection{Hydra}

The public release version of Hydra, upon which the code used 
in this work is based, has been extensively tested in a variety of 
astrophysical fields (Couchman 1991, Couchman, Thomas \& Pearce 1995, 
Thacker et al. 1999, Kay et al. 1999, Benson et al. 1999). It is an 
adaptive particle-particle, particle-mesh code that includes  
smoothed particle hydrodynamics (SPH) (Gingold \& Monaghan 1977; 
Lucy 1977) to follow the gaseous component. 

For this work we have chosen to use the latest public release version 
of Hydra, (Hydra3.0). This version incorporates a standard 
pairwise Monaghan type ${\bf r . v}$ viscosity rather than one based on 
the divergence of the local flow as used in earlier release versions of 
Hydra, because the ${\bf r . v}$ viscosity provides better 
shock capturing. We also employ the improved neighbour counting estimator 
as described by Thacker et al. (1999). To follow the radiative cooling of the 
gas we use the cooling rates calculated by Sutherland \& Dopita (1993) 
interpolated to the chosen gas metallicity.

In addition we have rewritten the SPH algorithm in order to perform complete 
neighbour counting even in low density regions. Previously, neighbour 
searching within Hydra was limited by the search length imposed by 
the size of the particle-particle grid, resulting in an uncomfortably  
high minimum resolved density. For our problem we wish to resolve 
the hot, low density gas within expanding supernovae shells 
and so have removed the maximum search-length restriction.

\subsection{Individual supernovae}

Here the mechanism for adding individual SNRs into a simulation is 
outlined.  

The initial stages of a SNRs evolution are complex as an enriched, 
very high temperature gas ploughs into the surrounding ISM.  We do not wish 
to simulate this phase of SNR evolution.  In order to deal with multiple 
SNRs over a large range of length scales we ignore evolution prior to the 
onset of the late stages of evolution characterised by the 
pressure-driven snowplough (PDS) phase (cf Cioffi et al. 1988).

The choice of the beginning of the PDS phase as the starting point of 
the evolution within the code is due to a number of considerations.  Firstly 
we wish to simulate the evolution of many SNRs in complex environments -  
the mass and time resolution (as well as the complex input physics) required 
to simulate the earlier stages of evolution are beyond our remit.  Secondly, 
the evolution of a SNR to the PDS phase is reasonably well understood 
and provides a good basis for our simulations, allowing us reduce the 
mass resolution required to model its evolution.  Lastly, 
the evolution of a SNR through the adiabatic Sedov phase even in 
a strong density contrast such as would be found in a molecular cloud 
is expected to be roughly spherical; significant non-spherical evolution 
would only be expected in the late PDS phase (Dopita \& Smith 
1986). In dense environments, such as molecular clouds, the radius 
at which the PDS phase starts is much 
less than 1 pc, thus for significant non-spherical evolution during the 
Sedov phase the density change must be on a similarly small scale.

The early evolution of SNRs comprise a short-lived initial free-expansion 
phase followed by an adiabatic expansion (the Sedov phase) once the 
mass of swept-up material exceeds the initial ejecta mass (Spitzer 
1978).  We will concern ourselves only with the later stages of SNR 
evolution after the Sedov phase has finished.

A few $\times 10^4$ yrs after the initial supernova, the temperature of the 
adiabatically expanding remnant falls to the point at which cooling 
is efficient in the outer regions: $T \approx (5-6) \times 10^5$ K.  At 
this point around half of the remnant mass forms a thin shell which expands 
into the ISM in a pressure-driven snowplough (McKee \& 
Ostriker 1977; Lozinskaya 1992) .  It is at the point at which the 
PDS phase begins that we will start to simulate the evolution of SNRs.

At the beginning of the PDS phase the thermal energy $E_T$ in the remnant 
is $E_T \approx 0.36 E_{51}$.  The temperature $T_{\rm PDS}$ of the 
interior gas is given by 

\begin{equation}
T_{\rm PDS} = 1.545 \times 10^{10} \frac{E_{51} {\rm pc}^3}{n_0 r_{\rm PDS}^3}  \,\,\,K,
\label{eqn:tpds}
\end{equation}

\noindent typically a few $\times 10^6$ K.  This leads to a 
value of the interior pressure in very good agreement with that of 
Chevalier (1974).  The thin shell (containing half the 
mass) is at around $5 \times 10^5$ K.

Initially a supernova inputs energy into the ISM such that the 
thermal energy is $E_T \approx 0.72 E_{\rm SN}$ and the remainder is in the
form of kinetic energy where $E_{\rm SN}$ is the total supernova 
energy, where the usual value is $E_{\rm SN} = 10^{51}$ ergs 
(Chevalier 1974).  By the time a remnant enters the PDS phase, around 50 per 
cent of the thermal energy 
will have been lost (Lozinskaya 1992).  The shell temperature is 
$(5 - 6) \times 10^5$ K and the interior pressure will have fallen to 
$P \approx E_{\rm SN}/(4 \pi R^3_{\rm PDS})$ where $r_{\rm PDS}$ is the 
radius at which the transfer to the PDS phase occurs (McKee \& Ostriker 1977).

The radius $r_{\rm PDS}$ at which the PDS phase starts is given by Cioffi 
et al. (1988) as  

\begin{equation}
r_{\rm PDS} = 14 E_{\rm SN}^{0.29} n_0^{-0.43} \zeta_m^{-0.143} {\rm pc}
\label{eqn:rpds}
\end{equation}

\noindent where $n_0$ is the ambient density in Hydrogen atoms cm$^{-3}$ 
and $\zeta_m$ is related to the metallicity (for Solar 
metallicity $\zeta_m = 1$).  

The initial velocity $v_{\rm PDS}$ of the shell is given by

\begin{equation}
v_{{\rm PDS}} = 413 n_0^{0.143} E_{51}^{0.071} \zeta_m^{0.214} {\rm \,\,\,km \,\,s}^{-1}
\label{eqn:vpds}
\end{equation}

These parameters (radius and velocity) are similar to those given by 
other authors (eg. Chevalier 1974; Falle 1981; Blondin et al. 1998).   
Differences are mainly due to the use of different cooling functions although 
the shell velocity predicted at the same radius is very similar for 
all calculations.

The main reasons for preferring the calculations of Cioffi et al. (1988) 
over other calculations are that they include a correction factor for 
non-Solar metallicities and their good agreement with the 
simulations of Thornton et al. (1998).

\subsection{SNRs in the code}

In the remainder of this section we compare our results for a SNR in a 
uniform medium with those of other authors as well as testing the 
convergence and self-consistency of our results.

SNRs are set-up by taking all of the particles in a sphere of radius 
$r_{\rm PDS}$ centred on the position of the supernova.  Half of the particles 
are unmoved but are heated up to $T_{\rm PDS}$ to represent the hot 
interior (uniform density and pressure, no bulk motions).  The other half of 
the particles are distributed on the surface of a sphere of 
radius $r_{\rm PDS}$ and given a velocity outwards from the centre of the 
sphere of $v_{\rm PDS}$ at a temperature of $5 \times 10^5$ K.

Low particle numbers within $r_{\rm PDS}$ have an effect on the evolution of 
the SNR due to shot noise causing significant non-sphericity.  Generally 
more than $\approx 60$ particles are required within $r_{\rm PDS}$ to model 
the SNR well although as few as 20 do a reasonable job (we would not 
wish however to trust simulations where the included particle number 
is this low).   

\subsubsection{SNR energy}

Table~\ref{tab:runs} gives the parameters used in 10 example runs 
covering a factor of 8 in $N$, 2.5 in box size and over 30 in mass 
resolution for the same physical initial conditions.  These runs are used in 
the next 2 subsections to illustrate the stability of the code at 
reproducing results over this range of parameters and to compare with 
other authors to show the ability of the code to well represent 
the post-Sedov phase evolution of a SNR.

\begin{figure}
\centerline{\psfig{figure=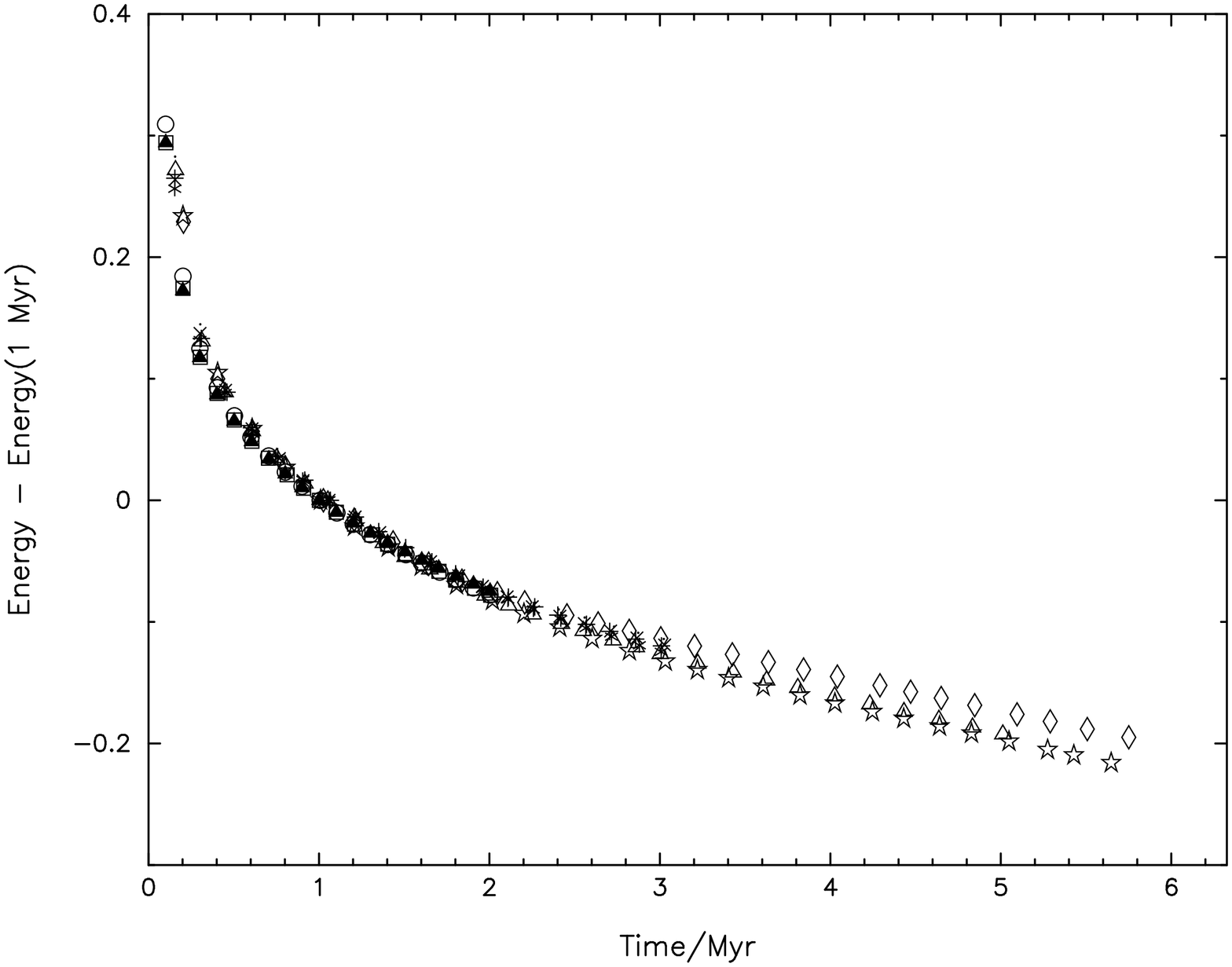,height=9.0cm,width=13.0cm,angle=270}}
\caption{The change in the total energy with time for the ten SNRs detailed  
in table~\ref{tab:runs} normalised to the total energy within the 
simulation volume at 1 Myr in units of $10^{51}$ ergs. The `lost' 
energy has been radiated away.}
\label{fig:alle}
\end{figure}

Figure~\ref{fig:alle} shows the change in the total energy within the 
simulation volume, normalised to the value at 1 Myr to take into account the  
different amounts of total thermal energy present in simulation volumes of 
varying size (in large boxes the thermal energy of the undisturbed gas 
is by far the largest contributor to the total energy).  The `lost' energy 
has been radiated away by the hot gas, the  energy conservation being 
good, a maximum of $0.1 - 0.2$ per cent. Good agreement 
is seen over a wide variety of particle mass and box size.  The small 
differences are mainly due to the requirement that there be 
an integer number of particles in the interior and shell of the SNR.  This 
leads to small differences between the initial thermal and kinetic energies 
of each model which amount to a few per cent, a difference which is amplified 
as time progresses.  For instance, the initial kinetic energy of runs 1019 and 
1020 differ by 4 per cent, by 6 Myr the difference in fig.~\ref{fig:alle} 
of $0.02E_{51}$ represents approximately 4 per cent of the total energy change.

\begin{figure}
\centerline{\psfig{figure=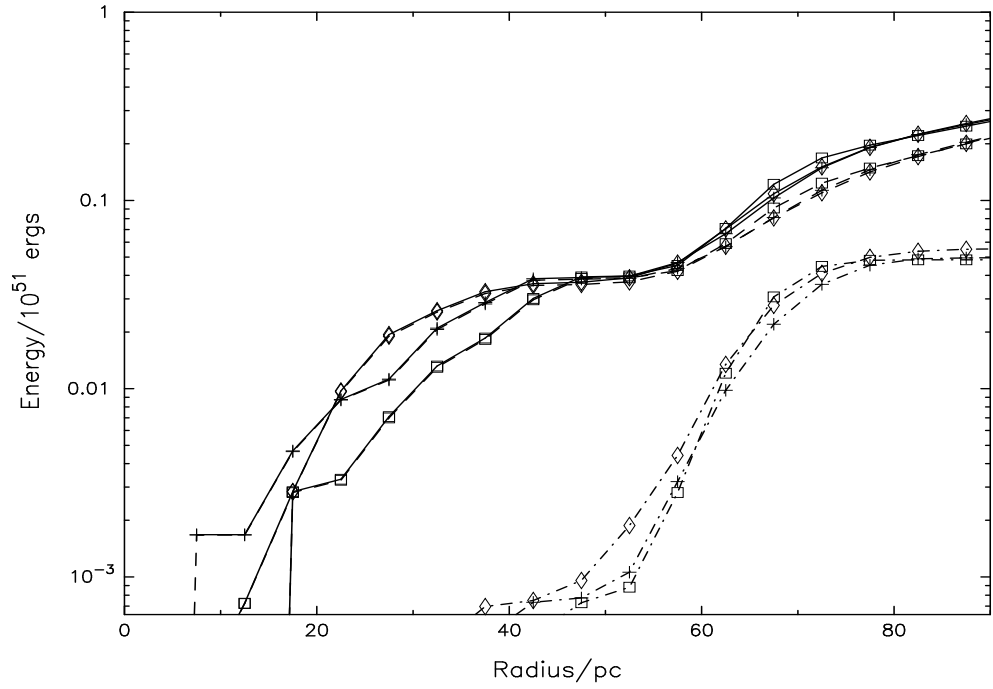,height=9.0cm,width=13.0cm,angle=270}}
\caption{The energy balance for the SNRs in runs 1002 (cross), 1017 
(square) and 1019 (diamond) at 1 Myr.  The solid lines show the 
sum of the thermal and kinetic energy, the dashed lines show  
the thermal and the dash-dot line the kinetic energy within each radius.}
\label{fig:energyshells}
\end{figure}

A more detailed examination of the energy balance for three of the 
SNRs is shown in fig.~\ref{fig:energyshells}, approximately 0.6 Myr after the 
supernova explosion. The solid lines show the sum of the 
thermal and kinetic energy within each radius with the dashed and 
dot-dashed lines showing the total thermal and kinetic energy respectively  
within each radius.  The outwardly moving shell is clearly visible.  The 
differences in interior thermal energy are due to shot noise and very low 
particle numbers in these regions - the high thermal energies of runs 1002 
and 1017 within the inner 10 pc are due to 1 hot particle. The agreement of
the energies over a factor of more than 30 in mass resolution is good and 
is representative of all runs.  The shell has a peak density at 
approximately 60 pc at 0.6 Myr and is wider where the mass resolution is
poor.  

\subsubsection{SNR expansion law}

The expansion of a SNR can be roughly characterised as a power law of the 
form $r \propto t^{\nu}$ (with $t$ being measured from the time of the 
supernova).  McKee \& Ostriker (1977) find for negligible external pressure 
that $\nu = 2/7\,(\approx 0.286)$.  Chevalier (1974) finds $\nu = 0.305$ for 
early times and for late evolution ($t > 0.75$ Myr) $\nu = 0.32$.  This 
is close to the Cioffi et al. (1988) offset solution where $r \propto 
(t - t_{\rm offset})^{0.3}$.   From 
inspection of fig.~3 in Thornton et al. (1998), they appear to obtain a value 
around $\nu = 0.3$.  All these values are below the power law for a 
purely momentum conserving snowplough where $\nu = 0.4$ (Spitzer 1978). 

\begin{figure}
\centerline{\psfig{figure=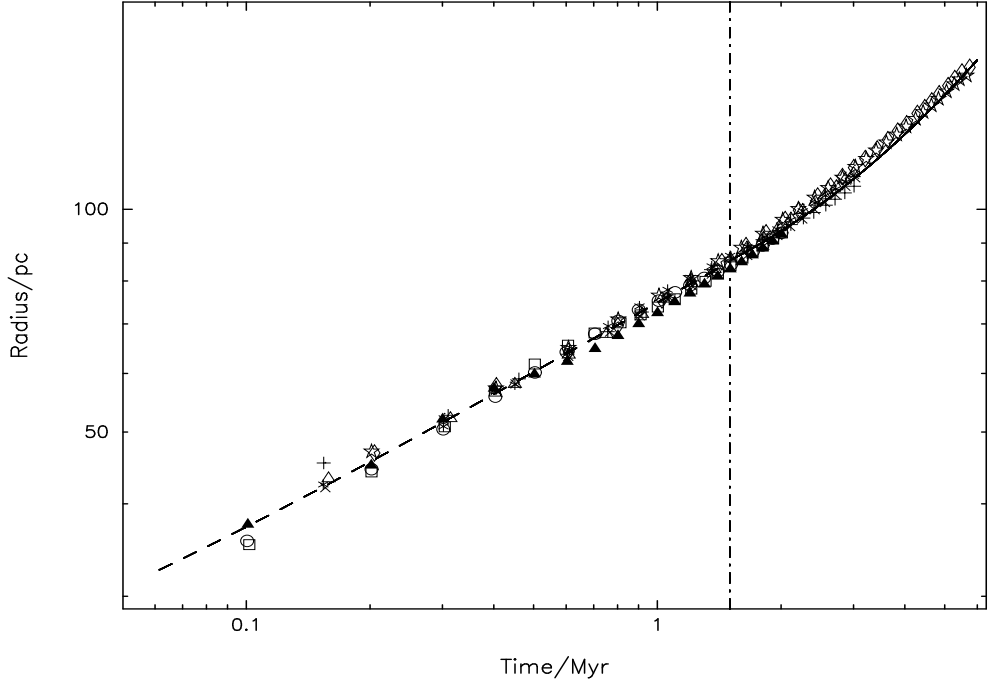,height=9.0cm,width=13.0cm,angle=270}}
\caption{The evolution of the shell radius with time for the ten
convergence runs detailed in table~\ref{tab:runs}.  The fits to different 
evolutionarystages are explained in the text.  The rough region of 
transition fromthe radiative pressure driven snowplough phase to the 
constant expansion phase is indicated by the dashed line at 1.5 Myr.}
\label{fig:revol}
\end{figure}

Figure~\ref{fig:revol} shows the evolution of the shell radius with 
time for the ten convergence test runs (which model the same physical 
conditions at different mass resolutions and box sizes) detailed in 
table~\ref{tab:runs}.  In fig.~\ref{fig:revol} the shell radius is 
determined by the mean radial distance from the centre of the SNR of 
the densest particles.  At low shell radii $r$ this is less reliable due to 
the smaller number of particles in the shock (especially when the 
mass resolution is poor).

The best fit to the first 1.5 Myr of evolution is shown by the dashed 
curve fitted with;

\begin{equation}
r(t) = r_{\rm PDS} \left( \frac{t}{t_{\rm PDS}} + 1 \right) ^{\nu}
\label{eqn:explaw}
\end{equation}

\noindent with $r$ in pc and $t$ in Myr ($t$ is the simulation time, ie. the 
time since the onset of the PDS phase), $t_{\rm PDS}$ is of order the 
duration of the PDS phase.  Leaving $r_{\rm PDS}$, $t_{\rm PDS}$ and $\nu$ 
as free parameters the best fit is given by $r_{\rm PDS} = 18.34$ pc, 
$t_{\rm PDS} = 0.012$ Myr and $\nu=0.317$.  For an $n_0 = 0.5$ 
cm$^{-3}$ ISM, $r_{\rm PDS}$ is actually 18.85 pc (from eqn.~\ref{eqn:rpds}) 
and $t_{\rm PDS} = r_{\rm PDS}/v_{\rm PDS}$ is around 0.02 Myr (Cioffi 
et al. 1988).

The late stages ($> 1.5$ Myr) are well fitted by the solid line, $r = 55.8 
+ 16.8t$, a reasonably good fit to free expansion of a sound wave, the 
sound speed in this case of 15.6 pc Myr$^{-1}$.

\subsubsection{Different ambient densities}

We have investigated the differences in SNR evolution within ISMs  
of different ambient densities.  Table~\ref{tab:sound} shows the range 
of ambient densities tested, covering 4 orders of magnitude.  Previous 
studies (with the exception of Thornton et al. 1998) have concentrated 
on investigating low ambient densities representative of the ISM in the 
Solar neighbourhood. 

\begin{figure}
\centerline{\psfig{figure=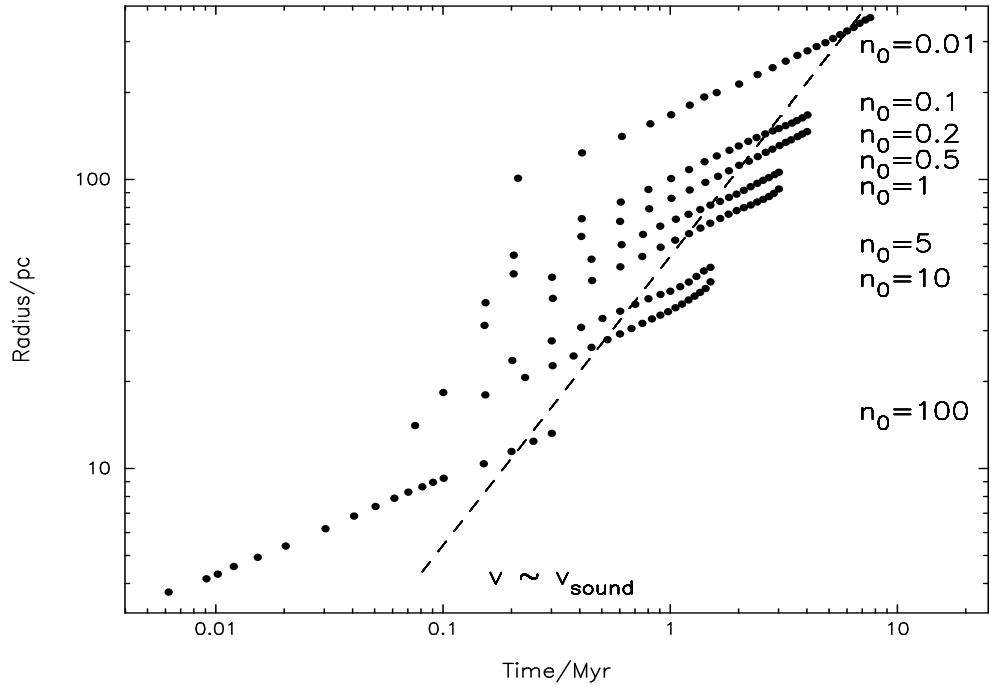,height=9.0cm,width=13.0cm,angle=270}}
\caption{The evolution of the shell radius for 8 different ambient 
density ISMs all at $10^4$ K detailed in table~\ref{tab:sound}.  
The hydrogen atom density per cm$^3$ is marked by each track.  The 
approximate time and radius at which the shell velocity falls to the ISM 
sound speed (and the tracks become curved) is marked by the dashed line.}
\label{fig:diffdens}
\end{figure}

Figure~\ref{fig:diffdens} shows the shell evolution for the range of 
densities presented in table~\ref{tab:sound}.  Also shown is the radius and 
time at which the shell velocity falls to the sound speed for each density.  

From the data presented in table~\ref{tab:sound} the radius at which
the sound speed is reached is fitted by a power-law in density of
$r_{\rm sound} = 65 n_0^{-0.37}$ pc and the time by a power-law of the
form $t_{\rm sound} = 1.2 n_0^{-0.38}$ Myr.  

These power-law fits to $r_{\rm sound}$ and $t_{\rm sound}$ are almost 
exactly what are expected.  Using eqn.~\ref{eqn:explaw} and the known 
dependencies on $n_0$ of $r_{\rm PDS} (\propto n_0^{-0.43}$ 
- eqn.~\ref{eqn:rpds}) and $t_{\rm PDS} (\propto r_{\rm PDS}/v_{\rm PDS} 
\propto n_0^{-0.57}$ -eqns.~\ref{eqn:rpds} \&~\ref{eqn:vpds}) gives 
for $t/t_{\rm PDS} >> 1$ 

\begin{equation}
r \propto n_0^{-0.25} t^{0.32}
\end{equation}

\noindent as $v_{\rm sound} = r_{\rm sound}/t_{\rm sound} = constant$ 
leads to the predictions that $t_{\rm sound} \propto n_0^{-0.36}$ 
and $r_{\rm sound} \propto n_0^{-0.37}$, very close to the fitted 
relations from the simulations.

\begin{figure}
\centerline{\psfig{figure=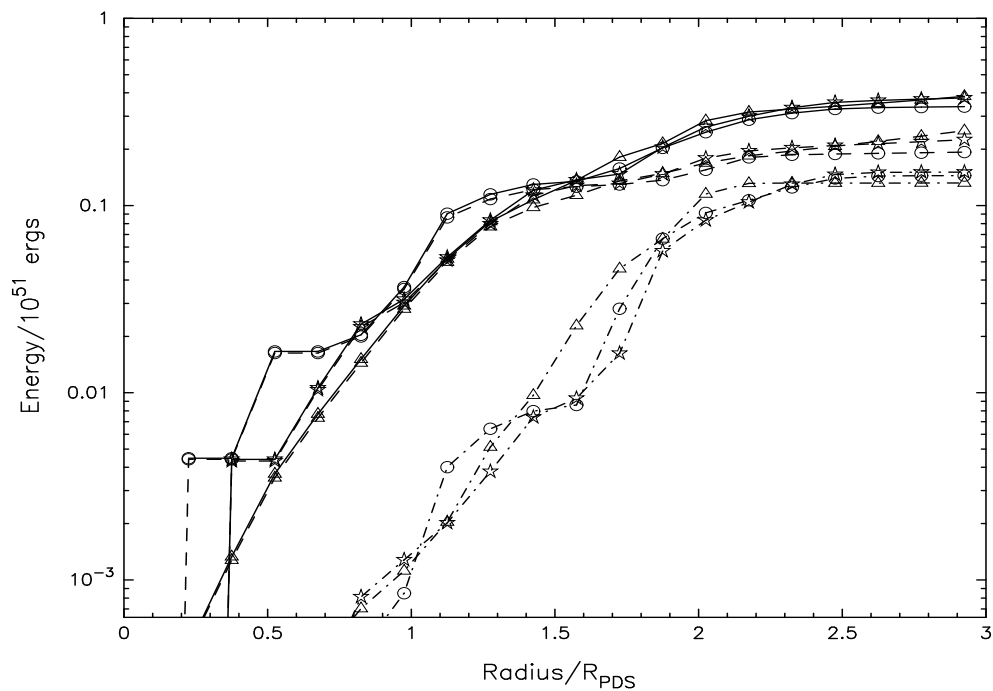,height=9.0cm,width=13.0cm,angle=270}}
\caption{The total (solid line), thermal (dashed line) and kinetic (dot-dash 
line) energies within each radius at a shell radius of $\approx 2 
\times r_{\rm PDS}$  the radial distribution being given in terms of 
$r/r_{\rm PDS}$ for three different densities of $n_0 = 0.01$ cm$^{-3}$ 
(triangles), $n_0 = 0.5$ cm$^{-3}$ (stars) and $n_0 = 100$ cm$^{-3}$ (circles).}
\label{fig:scalepds}
\end{figure}

The initial energy (thermal and kinetic) of the SNR and the evolution of 
the total energy is approximately the same for each density if the evolution 
of a SNR through the PDS phase is scaled so that the radial properties 
are in units of $r_{\rm PDS}$ as illustrated in fig.~\ref{fig:scalepds}.  
The radial total, thermal and kinetic energies are plotted against 
$r/r_{\rm PDS}$ for 3 models spanning a range of 4 orders of magnitude 
in density for the time when $r_{\rm shell} \approx 2 \times r_{\rm PDS}$.  
For $n_0 = 0.01$ cm$^{-3}$, $2r_{\rm PDS} \approx 200$ pc, for $n_0 = 0.5$ cm$^{-3}$, $2r_{\rm PDS} \approx 38$ pc while for $n_0 = 100$ cm$^{-3}$, 
$2r_{\rm PDS} \approx 3.9$ pc. 

As is clear in fig.~\ref{fig:scalepds} the energy profiles of the SNRs in 
terms of $r/r_{\rm PDS}$ are very similar, indeed the main differences are 
due to the fact that the output radii shown are not exactly 
$r_{\rm shell} = 2 \times r_{\rm PDS}$ but differ by a few per cent (due to
the finite number of outputs).

When $r = 2 \times r_{\rm PDS}$ the mass of the hot interior is around twice  
the interior mass at the onset of the PDS phase as most of the swept-up 
material remains in the dense shell.  The interior temperature falls 
by a factor of around 2.7 while 
the shell temperature has fallen to $\approx 10^4$ K.  The shell velocity 
has fallen by a factor of 5 (as expected if $v$ scales as 
$v_{\rm PDS}(t/t_{\rm PDS} + 1)^{-0.68}$, cf. eqn.~\ref{eqn:explaw}).  The 
scaling of the SNR properties with $r_{\rm PDS}$ across all densities is 
because the energy content of SNRs is density independent - a doubling of 
the volume of the SNR will result in the same energy changes throughout 
the SNR independent of the ambient density (as swept-up mass dominates).

\subsubsection{Late-starting SNRs}

\begin{figure}
\centerline{\psfig{figure=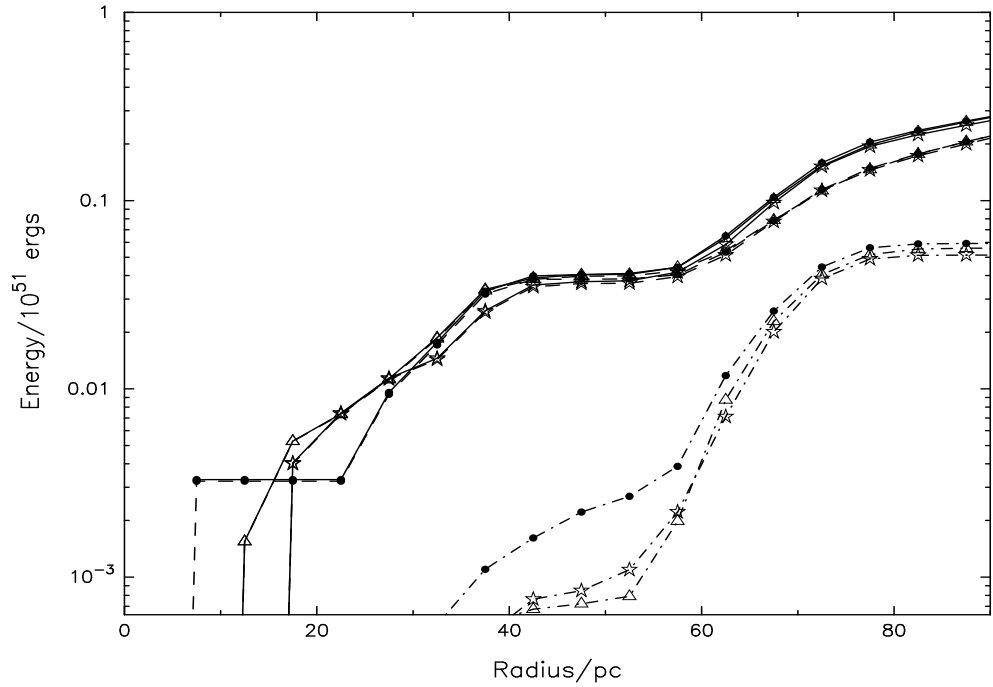,height=9.0cm,width=13.0cm,angle=270}}
\caption{The radial energy profiles of runs 1003 (stars) and 1018 
(triangles) at 1 Myr  compared to that of a run started at $r = 2 \times 
r_{\rm PDS}$ (circles).  The shell radius at this time is $\approx 
55$ pc.  The energies shown are the total (full line), thermal (dashed line) 
and kinetic (dot-dash line) within each radius.}
\label{fig:snrcomp}
\end{figure}

Figure~\ref{fig:snrcomp} shows the late evolution of three SNRs, 
two started at the onset of the PDS phase and the other started at $r = 2 
\times r_{\rm PDS}$ with $T_{\rm shell}=10^4$ K and the interior 
temperature given by  $T_{\rm SNR}=0.37 \times 
T_{\rm PDS}$ (where $T_{\rm PDS}$ is given by eqn.~\ref{eqn:tpds}) and 
$v_{\rm shell} = 0.2 v_{\rm PDS}$.  The total energy within the 
late-starting SNR is very similar to that 
in those that started at $r_{\rm PDS}$ but the relative contributions of the 
kinetic and thermal energies within the bubble are slightly different.  
This is due to the shorter amount of time which the $2r_{\rm PDS}$ simulation 
has had to convert bulk motions into thermal energy within the bubble.  Again 
the main differences between the simulations are due to outputs not 
being at exactly the right times.   The ability to start modelling each SNR 
at $2 \times r_{\rm PDS}$ will become very useful when we wish to 
add SNRs to massive gas clouds reducing the required mass resolutions 
in simulations by a factor of 8 (see section 3).

\subsubsection{Metallicity effects}

The effect of different metallicity environments can be significant 
during the early evolution of SNRs, during the PDS phase however the 
differences due to metallicity are found to be negligible (see also Thornton 
et al. 1998).  Initially a metallicity of $\zeta_m = 0.01$ will make a factor
of 1.93 difference in $r_{\rm PDS}$ and 0.37 in $v_{\rm PDS}$ as compared to 
a $\zeta_m = 1$ SNR (from eqns.~2 \& 5 respectively).  However, the effect of 
these different initial SNR conditions is not as significant as it might 
first appear, because the velocity of a $\zeta_m = 1$ SNR at $r = 1.93 \times 
r_{\rm PDS}(\zeta_m=1)$ is $\approx 0.22 v_{\rm PDS}(\zeta_m=1)$ (see 
above) resulting in only slightly different expansions.

The subsequent evolution of SNRs with different metallicities through the PDS 
phase are virtually indistinguishable (as also found by Thornton et al. 
1998).  The evolution of the shell radii is slightly different at early 
times due to different values of $r_{\rm PDS}$ but as mentioned above this 
difference is soon nearly cancelled out and at late times the shell radii 
differ by only a few pc.  When SNRs are started with the same $r_{\rm PDS}$ 
but different metallicities the difference is negligible.  The evolution 
of the total energies within each SNR are also remarkably similar.  Differences 
in metallicity affect the cooling of the hot interior and hence the 
pressure and the extra driving force on the shell. However, during 
expansion into a uniform ambient medium the shock acceleration is dominated 
by the deceleration due to the sweeping-up of material, making metallicity 
(which effects the pressure-driven acceleration) a minor factor in the 
late-time evolution of the SNR.

The differences in energy between different metallicities are accounted for 
by the different initial conditions at the onset of the PDS phase 
however it is not clear that even the early stages of SNR evolution would 
be affected by low-metallicity effects.  A type II supernova with a 
progenitor mass of $25 M_{\odot}$ will produce around $2.4 M_{\odot}$ of heavy 
metals (Tsujimoto et al. 1995), the majority ($1.8 M_{\odot}$) being O$^{16}$.  In 
the case of a SNR expanding into an ISM of density $n_0 = 100$ cm$^{-3}$ 
(where the mass within $r_{\rm PDS}$ is $\approx 75.5 M_{\odot}$) this will enrich 
the swept-up material from $\zeta_m = 1$ (Solar) to $\zeta_m = 2.6$ and 
from $\zeta_m = 0.01$ to $\zeta_m = 1.6$.  Such enrichment is very significant and will alter the evolution of the pre-PDS phases.  However, by 
$5 r_{\rm PDS}$ the amount of swept-up material is $9440 M_{\odot}$ and has 
been enriched from $\zeta_m = 0.01$ to $\zeta_m = 0.023$, a fairly 
insignificant (from the point of view of cooling) amount.  Thornton et al. 
(1998) did not find a significant effect when including the metals 
ejected from the supernova (Thornton 1999).  Obviously this enrichment and 
its history is of vital importance for understanding chemical evolution, and 
the enrichment history can be followed by this code (once a number 
of assumptions about mixing have been included), but it has little bearing 
on the late-time evolution of an individual SNR in a uniform medium.   

\subsubsection{Computational aspects}

The softening included in the simulations does not greatly affect
the results of the shell expansion and energy transfer.  Similar runs 
differing in softening are virtually indistinguishable in their results (two 
such runs, differing by a factor of over 4 in softening - 1002 \&
1005 - are included in fig.~\ref{fig:alle}).  The softening selected for 
simulations should be as small as is reasonably practicable given 
that the smallest region in which results can be believed in detail is 
no less than the softening length, which also determines the width of
any shock fronts that are present. Unfortunately the number of
timesteps required rises as the softening length is decreased.

The width of the shell that forms the snowplough is difficult to
determine accurately as it becomes slightly aspherical as it evolves
due to initial Poisson fluctuations in the particle positions. This
effect is worse for the models which began with only a few particles
in the SNR. When spherically averaged profiles are produced this leads
to an artificial broadening of the snowplough and general smoothing of
the steep shock fronts.  The snowplough width is expected to be a few
pc (Thornton et al. 1998), in these simulations the shock width is 
approximately the softening.

\subsection{Summary}

In this section we have shown that the code is stable and converges for 
reasonable selections of box size and particle mass.  The code is able to 
reproduce the results of other authors for the situation of a SNR 
expanding into a uniform ambient medium during the PDS phase of 
evolution.  In the next section we will place SNRs in the centres of 
Plummer model gas clouds.

\section{CENTRAL SUPERNOVAE IN A GAS CLOUD}

While the evolution of a SNR in a uniform density ISM provides a good test 
of the ability of the code to reproduce the characteristics of a SNR 
as found in more detailed hydrodynamic calculations, it is not 
physically realistic.  The effects of a supernovae in a stratified medium 
(eg. a gas cloud) is a more practical problem.  Massive stars are found in 
young clusters, their 
lifetimes being so short that they are expected to still be embedded in 
gas remaining from star formation.  The 
significant evolution of a young SNR is therefore expected to occur 
in the confines of its parent GMC.  Massive stars are also found to be 
centrally-concentrated in clusters (Hillenbrand 1997; Carpenter et al. 1997) 
so that an investigation of a central SNR is a good approximation to the 
evolution of the first SNR in a cluster.

Some analytic calculations of the effects of supernovae in a gas cloud 
have been made by Dopita \& Smith (1986) and Morgan \& Lake (1989) to 
estimate the number of supernovae required to totally disrupt a gas 
cloud of a given mass.  Morgan \& Lake (1989) using more detailed 
cooling functions than Dopita \& Smith (1986) found that the 
minimum mass of a ($1/r^2$) cloud for it to confine a single central 
supernova was $4 \times 10^4 M_{\odot}$.

As noted in section~2.3, Dopita \& Smith (1986) find that a SNR within the 
high-density environment of a gas cloud will remain roughly spherical 
during the adiabatic phase of its evolution.  Using this result we 
may place SNRs in the cloud at the start of the PDS phase in the 
same way as they are placed in the uniform medium simulations.  The high 
densities in gas clouds ($n_0 > 10^4$ cm$^{-3}$) means that the 
adiabatic phase will end whilst the shell radius is very small ($\ll$ pc).

Cloud initial conditions are based on a Plummer model ($n=5$ polytrope) 
with mass distribution;

\begin{equation}
M(r) = \frac{Mr^3}{R^3} \frac{1}{[1 + r^2/R^2]^{3/2}}
\end{equation}

\noindent where $M$ is the total mass and $R$ is the scale length.  The 
half-mass radius of a cloud is then $\approx 1.3R$.  Clouds are constructed 
out to a maximum radius $r_{\rm max}$ of a few $R$ (normally $r_{\rm max}=20$ 
pc), resulting in the actual mass being slightly less than $M$.

In this paper we will only deal with Plummer model clouds.  Obviously 
the effects of a SNR will depend, at least to some extent and possibly 
very significantly, upon the density distribution of the parent 
cloud.  These will be dealt with in detail in a paper to follow.

Observationally, Giant Molecular Clouds (GMCs) can approximated very 
roughly by clouds with $M =$ a few $\times 10^4$ to a few $\times 10^5 
M_{\odot}$ and $R =$ a few pc (eg. Harris \& Pudritz 1994) which we used as 
initial conditions for our simulations (full details in 
table~\ref{tab:models}). Before inserting any 
supernovae, our clouds are allowed to relax to a stable state in which 
they are virialised.  Support is provided by the thermal 
energy of the gas and the bulk kinetic energy is negligible (although the 
temperature of the gas represents the turbulent velocity which supports the 
cloud rather than the molecular temperature).  The clouds are self-gravitating 
as are real giant molecular clouds (Rivolo \& Solomon 1988) and so require no 
pressure confinement from an external hot diffuse gas.  

The insertion of a supernova has the effect of adding considerable amounts 
of kinetic and thermal energy to the cloud.  By the end of the Sedov 
phase, the extra energy amounts to around $6.4 \times 10^{50}$ ergs, 
comparable to or greater than the potential energy of the system 
(roughly $8.6 \times 10^{40} (M/M_{\odot})^2/(R/{\rm pc})$ ergs).  The shock from 
the supernova then passes through the the cloud.

\subsection{Some numerical considerations}

The number of particles within $r_{\rm PDS}$ required for the SNR evolution 
to be a good approximation to more complex simulations is $> 40$ 
with $> 60$ giving the best results (see section 2).  For a Plummer 
model cloud the central density is

\begin{equation}
\rho_0 = \frac{3 M}{4 \pi R^3} 
\end{equation}

\noindent where $M$ is the mass of the cloud and $R$ is the Plummer 
scale length.  Using 
equation \ref{eqn:rpds} the mass interior to $r_{\rm PDS}$ at the 
cloud core will be

\begin{equation}
M_{\rm PDS} \approx 5.48 \frac{R_{\rm pc}^{6/7}}{M_5^{2/7}} \,\,\, M_{\odot}
\end{equation}

\noindent where $R_{\rm pc} = R/$pc and $M_5 = M/10^5 M_{\odot}$ (we shall be 
using $R_{\rm pc}$ and $M_5$ throughout the rest of this section).  We require 
a mass resolution at least 60 times smaller than this giving, for 
$R_{\rm pc} = 3.5$ and $M_5=1$ a minimum particle number of 
$N > 3.74 \times 10^5$.

Such resolutions, while attainable, require significant computing time 
to run.  Our aim here is to introduce SNRs in such a way as to allow 
many simulations covering a variety of initial conditions to be
completed, so as to explore the parameter space.  Starting SNR evolution 
at $r = 2 \times r_{\rm PDS}$ (see Section 2.3.4) allows a factor of 8 
reduction in the mass resolution and therefore the number of particles 
required to model the cloud.  In the case of a $10^5 M_{\odot}$ cloud $2 
\times 32^3$ 
particles are required, on a workstation each timestep takes around 3 
minutes resulting in a 20 Myr simulation in around 30 hours.

\subsection{Feedback as a function of cloud structure}

Here we present the analysis of the energy and mass feedback as 
a function of $M_5$ and $R_{\rm pc}$, the cloud models are 
presented in table~\ref{tab:models}.  Initially 
all clouds are virialised and particles are said to be lost once they 
cross the box edges which are a distance 
$2 r_{\rm max}$ from the centre of the cloud (at a distance roughly 
equal to the tidal radius of a GMC).  It is these lost particles 
that comprise the feedback from the cloud into the larger ISM.  In most 
cases the SNRs are started at $2 r_{\rm PDS}$.

As a 'standard' model we use the run with parameters $M_5 = 1$, 
$R_{\rm pc} = 3.5$ to illustrate the general features of SNR evolution 
in a Plummer cloud.  Initially the cloud has $0.0356 \times 10^{51}$ ergs 
of thermal energy and a potential energy of $-0.0711 \times 10^{51}$ 
ergs.  The SNR starting at $2 r_{\rm PDS}$ adds a total of $0.123 
\times 10^{51}$ ergs of kinetic energy and $0.289 \times 10^{51}$ ergs 
of thermal energy (calculated from section 2) creating a net positive 
energy for the system of $0.38 \times 10^{51}$ ergs.

\subsubsection{The evolution of the SNR}

\begin{figure}
\centerline{\psfig{figure=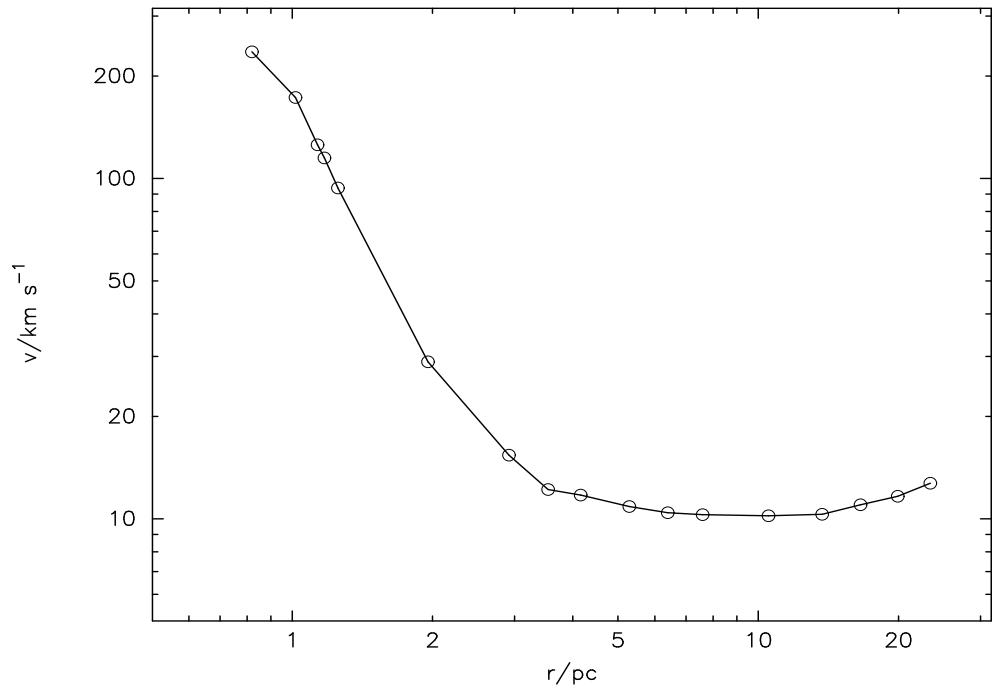,height=9.0cm,width=13.0cm,angle=270}}
\caption{The evolution of the shock velocity with radius for 
the standard $M_5 = 1$, $R=3.5$ cloud.}
\label{fig:vr5000}
\end{figure}

The evolution of the shell velocity with radius for the standard model is 
shown in fig.~\ref{fig:vr5000}.  It is typical of all runs that the 
velocity drops from its initial value ($v_{\rm PDS}$ or $0.2 v_{\rm PDS}$ 
of started at $2 r_{\rm PDS}$) reaching a minimum value $v_{\rm min}$ at around 
2 to 3 $R_{\rm pc}$ before accelerating to around $1.5 \times v_{\rm min}$ 
on leaving the box.

It is found later that $v_{\rm min}$ and associated quantities may be 
used to parameterise all of the effects of a SNR on a gas cloud.  In this 
subsection we sketch a model of the SNR evolution that predicts $v_{\rm min}$ 
in terms of the two cloud parameters $M_5$ and $R_{\rm pc}$.

The velocity reaches $v_{\rm min}$ when the acceleration due to the 
interior pressure matches the deceleration due to the sweeping-up of 
material.  The force on the shell due to pressure $F_{\rm press}(r)$ goes as 

\begin{equation}
F_{\rm press} = (4 \pi r^2) P
\label{eqn:fpress}
\end{equation}

\noindent where $P$ is the pressure when the shell radius is $r(t)$.  

The initial radius of the shell ($r_{\rm PDS}$) is very small compared 
to $R$ and as adiabatic cooling goes as $(r_{\rm PDS}/r)^5$ the 
majority of the cooling will be adiabatic, especially at late times so 

\begin{equation}
P \approx P_{\rm PDS} \frac{r_{\rm PDS}^5}{r^5} 
\label{eqn:adia}
\end{equation}

\noindent with $P_0$ given by $\rho_0 k T_{\rm PDS}$ (where $\rho_0$ is the 
central density $\propto M_5/R_{\rm pc}^3$).  The initial temperature 
$T_{\rm PDS}$ is given by eqn.~\ref{eqn:tpds} as 

\begin{equation}
T_{\rm PDS} \propto \frac{M_5^{0.29}}{R_{\rm pc}^{0.87}}
\label{eqn:t0}
\end{equation}

The force due to the sweeping-up of material (momentum conservation) 
$F_{\rm mom}(r)$ is

\begin{equation}
F_{\rm mom}(r) = (4 \pi r^2) \rho(r) v(r)^2
\label{eqn:fmom}
\end{equation}

\noindent equating eqns.~\ref{eqn:fpress} and~\ref{eqn:fmom} using 
eqn.~\ref{eqn:adia} gives

\begin{equation}
\rho(r_{\rm min}) v_{\rm min}^2 = P_0 \frac{r_{\rm PDS}^5}{r_{\rm min}^5}
\label{eqn:equal}
\end{equation}

\noindent using eqn.~\ref{eqn:t0} and assuming that $r_{\rm min} \gg R$ 
gives 

\begin{equation}
v_{\rm min} \propto \frac{R_{\rm pc}^{0.29}}{M_5^{0.93}}
\label{eqn:vsimp}
\end{equation}

\noindent Fitting the powers of $M_5$ and $R_{\rm pc}$ simultaneously 
gives a best linear fit to $v_{\rm min}$ of 

\begin{equation}
v_{\rm min} = 6.58 \frac{R_{\rm pc}^{0.32}}{M_5^{0.77}}  \,\,\,\,{\rm km}\,\,\,{\rm s}^{-1}
\label{eqn:vmin}
\end{equation}

\noindent which is compared to the simulations in 
fig.~\ref{fig:vminMR}.  The similarities between eqns.~\ref{eqn:vsimp} 
and~\ref{eqn:vmin} are remarkable considering the very simple assumptions 
that went into the formulation of eqn.~\ref{eqn:vsimp}.

The accelerating force due to the pressure goes as $r^{-5}$ and so 
rapidly becomes negligible at high $r$.  It is still able, however, 
to increase the velocity of the shell by the time it leaves the cloud 
to $\approx 1.5 v_{\rm min}$.

The above rather simplistic treatment avoids (as is necessary) many 
details of the actual SNR evolution in a cloud.  In the simulations 
the main shell of the SNR is followed by a weaker shock which is 
created by the shocking of infalling material filling the interior 
when it reaches the centre (and meets other infalling material).  This 
creates a very complex density/pressure structure in the cloud as 
the SNR evolves.  

We know go on to show how the effects of a SNR on a cloud can be 
parameterised very simply in terms of $v_{\rm min}$ and closely related 
quantities.

\subsubsection{Metallicity effects}

As the temperature factor $T_0$ is so important in setting the value 
of $v_{\rm min}$, then metallicity, $Z$, is an important factor in that in 
clouds of lower $Z$ than the fiducial Solar $Z=1$, cooling will be less 
efficient and $T_0$ and $v_{\rm min}$ higher. In the case of a 
$Z=0.01$ metallicity cloud the kinetic energy lost 
increases to $2.5 \times 10^{49}$ ergs and the mass lost to $5.4 \times 
10^4 M_{\odot}$, a 67 per cent increase on the $Z=1$ cloud in both 
cases.  For a $Z=0.1$ cloud the increase in both values is 46 per 
cent. This would indicate that the effects of the first generation
of supernovae would have been much more dramatic than those which occurred
later once metal enrichment had taken place.

\subsubsection{Mass loss and disruption}

The mass loss from the cloud would be expected to be related the escape 
velocity from the cloud at the point where the minimum velocity is 
reached.  The escape velocity at $2R_{\rm pc}$ is given by 
$18.6 \sqrt{M_5/R_{\rm pc}}$ km s$^{-1}$ and it is found that if 
$v_{\rm min} > 
v_{\rm esc}(r_{\rm min})$ then the cloud completely disrupts, ie. all of 
the mass is lost.  At the other extreme if  $v_{\rm min} < 0.2
v_{\rm esc}(r_{\rm min})$ then no mass loss (and hence no feedback) 
occurs.  In the intermediate regime the mass loss is related to 
$v_{\rm esc}(r_{\rm min})$ by

\begin{equation}
\frac{M_{\rm lost}}{M_5} = 0.25 \times \left( \frac{v_{\rm min}}{v_{\rm esc}} \right) ^{2.5}
\label{eqn:mlost}
\end{equation}

\noindent as illustrated in fig.~\ref{fig:mlost}.

\subsubsection{Energy feedback}

The amount of energy returned to the ISM from a cloud would be 
expected to be related to the kinetic energy of the cloud at 
$v_{\rm min}$ given by $T_{\rm min} = 1/2 M_{\rm shell} v_{\rm min}^2$ 
where $M_{\rm shell}$ is the mass of the shell at $r_{\rm min}$, most  
simply we might expect $M_{\rm shell} \propto M_5$.

Illustrated in fig.~\ref{fig:elost} is the feedback energy 
from table~\ref{tab:models} against $T_{\rm min}/10^{51}$ ergs $= 
0.001 M_5 (v_{\rm min}/$km s$^{-1})^2$.  Two linear 
regimes are obvious in the behaviour of the feedback energy.  In 
the upper limit where the cloud has totally disrupted the 
feedback energy is fitted by 

\begin{equation}
E_{\rm lost} = -0.0012 + 0.25 M_5 v_{\rm min}^2 \,\,\,\, \times 
10^{51}\,\,\,{\rm ergs}
\label{eqn:elost1}
\end{equation}

\noindent in the lower limit where the mass loss is negligible the 
feedback energy is 

\begin{equation}
E_{\rm elost} = -0.012 + 0.25 M_5 v_{\rm min}^2 \,\,\,\, \times 
10^{51}\,\,\,{\rm ergs}
\label{eqn:elost2}
\end{equation}

\noindent as can be easily seen the slope of these two relationships is 
the same and the constant factor, different by exactly one order of 
magnitude, is the only difference.  The intermediate stage marks the 
change from the total cloud destruction and high feedback to the low 
mass loss and low feedback regime.  This region is only a very small 
region of the total $M_5 - R_{\rm pc}$ parameter space.

\subsubsection{The final state of clouds}

As stated above when $v_{\rm min} > v_{\rm esc}$ a single SNR is enough 
to disrupt a cloud completely and when $v_{\rm min} < 0.2 v_{\rm esc}$ 
no mass loss occurs at all.  In the intermediate regime some 
gas is retained in a bound object - a new cloud with lower mass 
than the original.

The ratio of the final to initial potential energy, $\Omega_f/\Omega_i$ is 
related again to the ratio of the minimum velocity of the shell and 
the escape velocity $v_{\rm min}/v_{\rm esc}$

\begin{equation}
\frac{\Omega_f}{\Omega_i} = 1.26 - 1.05 \frac{v_{\rm min}}{v_{\rm esc}} 
\label{eqn:potlost}
\end{equation}

\noindent as illustrated in fig.~\ref{fig:potlost}.  This relationship is 
only valid for $0.3 < v_{\rm min}/v_{\rm esc} < 1$, beyond 1 the cloud is 
totally destroyed and $\Omega_f/\Omega_i = 0$ and less than 0.3 
$\Omega_f/\Omega_i$ asymptotes to 0.

The change in potential energy of a gas cloud due to a central supernova 
can sometimes be significantly more important than the feedback of 
kinetic energy as illustrated in table~\ref{tab:models}.  Clouds lose 
all of their initial potential energy if they are completely disrupted.  If 
the cloud loses only part of its mass then the significant energy change 
occurs as a loss of potential energy from the cloud (a net gain of 
energy).  Even when no mass loss or feedback occurs a small 
change (of the order of a few percent) in the potential energy is observed 
as the cloud expands slightly.

As fig.~\ref{fig:final} shows, the remaining material settles back into 
a new configuration that can also be described as a Plummer model 
with a lower central density and larger scale length.  Thus knowing the 
initial mass $M$ and characteristic radius $R$ gives the final 
characteristic radius $R_f$ via;

\begin{equation}
R_f = \left( 1 - \frac{M_{\rm lost}}{M} \right) ^2 R \left( \frac{\Omega_f}{\Omega_i} \right) ^{-1}
\end{equation} 

\noindent Knowing the final mass and characteristic radius of the 
cloud allows the effects of further central supernovae on that cloud to  
be calculated - as long as the interval between supernovae is greater 
than the time required for the cloud to relax to a new equilibrium.

The time taken for a $5 \times 10^5 M_{\odot}$ cloud to recover from 
a central supernova is very short, the majority of the SNR's kinetic 
and thermal energy is radiated away in $< 1$Myr.  Assuming a Salpeter 
IMF, the number of massive stars within a cloud that will go supernovae 
is $N_{\rm SNe} \approx 0.006 M \epsilon$ where $M$ is the mass of the 
cloud and $\epsilon$ is the star formation efficiency.  Taking 
$\epsilon = 0.01$ and $M=5 \times 10^5 M_{\odot}$ gives $N_{\rm SNe} = 30$. If 
these supernovae occur centrally and evenly spaced over a 30 Myr 
period then no ejecta will escape the cloud and the cloud will not 
be disrupted by the SNRs.

These results hold for Plummer model clouds that can be characterised 
solely by $M$ and $R$.  This study will be extended to other density 
distributions in a future paper which we are preparing.  

\section{CONCLUSIONS}

The Hydra $N$-body SPH code has been extended to allow the 
simulation of the evolution of supernova remnants (SNRs) from the onset of 
the pressure-driven snowplough (PDS) phase. In section 2 this code was seen 
to be able to produce convergent results over a wide range of parameter 
space and reproduce the results on the evolution of SNRs from a variety 
of previous authors.  The power of this code is the ability to simulate 
the evolution of SNRs in a variety of environments using a workstation in 
a reasonable time (of the order of days).

This code represents the first time that we are able to model multiple, 
interacting SNRs in gas clouds and complexes.  This paper is the 
first detailed analysis of SNR evolution in a gas cloud to examine the 
feedback parameters so important in galaxy formation and evolution 
calculations. 

In section 3 we presented new results on the effect of a single, central 
supernova on Plummer model gas clouds of various masses and characteristic 
radii.  The results of Dopita \& Smith (1986) and Morgan \& Lake (1989) 
were found to be too simplistic.  The evolution of a central SNR is 
very complex, the late stages of evolution are governed by pressure-driving 
from a hot interior accelerating the SNR shell down the density 
gradient.  

Feedback from a cloud is defined to be the mass and energy 
which pass out of our simulation box (whose size is approximately the 
tidal radius of the cloud).  The results may be summarised as:

\begin{itemize}

\item{The efficiency of energy feedback, mass loss and cloud destruction for 
a central supernova in a Plummer cloud of mass $M$ and characteristic 
radius $R$ is related to the minimum velocity $v_{\rm min}$ that the 
shock reaches during it's evolution.  This minimum velocity is given by 
$v_{\rm min} = R_{\rm pc}^{0.32}/M_5^{0.77}$.}

\item{For $v_{\rm min} > v_{\rm esc}$ the cloud is totally destroyed, while 
for $v_{\rm min} < 0.2 v_{\rm esc}$ the SNR is completely contained and no 
feedback occurs.}

\item{The mass lost from a cloud is related to the ratio of the minimum 
velocity to the escape velocity $v_{\rm min}/v_{\rm esc}$ as 
$M_{\rm lost}/M = 0.25 (v_{\rm min}/v_{\rm esc})^{2.5}$.} 

\item{The energy feedback $E_{\rm lost}$ has two main regimes with equal 
slopes where $E_{\rm lost} = -C_E + 0.25M_5V_{\rm min}^2 \times 10^{51}$ 
ergs where $C_E$ = 0.0012 when $v_{\rm min}/v_{\rm esc} > 1.1$ and 
$C_E$ = 0.012 when $v_{\rm min}/v_{\rm esc} < 0.9$.}

\item{The loss of (negative) potential energy is often the largest 
(positive) increase of energy in the system and at least of order the 
feedback of kinetic energy, the feedback of thermal energy being negligible 
in comparison to both.}

\item{The final state of a cloud that is not destroyed is close to a Plummer 
model with final characteristic radius $R_f$ related to the initial
parameters by  $R_{f} = ( 1 - M_{\rm lost}/M ) ^2 R ( \Omega_f/\Omega_i) ^{-1}$.}

\item{The efficiency of feedback increases rapidly with decreasing 
metallicity suggesting that feedback at early epochs was far more 
efficient.}

\end{itemize}

Our simulations are obviously not perfect.  The code takes no account of 
magnetic fields (following standard astrophysical practice) which may well be important, especially at late times.  In addition the feedback from massive 
stars into the cloud before they become supernovae is neglected, although 
in the dense environments of cloud cores we may be justified in ignoring 
this effect (Franco, Garc\'{i}a-Segura \& Plewa 1996).  A simple method 
of including such effects in the code is being developed.  Despite this 
we believe that this code represents a  significant step in modelling 
the effects of multiple SNRs on clouds and the larger ISM.

\subsection{Future directions}

As noted in the introduction, a code such as this has many interesting
applications.  It can be used to investigate how gas clouds of
different sizes and shapes are affected by internal supernovae, calculating 
the feedback of energy and mass into the ICM.  We are in the process of 
preparing papers that will expand the current work into an investigation 
of off-centre SNRs in gas clouds and of the effect of different density 
distributions on feedback parameters.

In addition we will be able to investigate how the cloud is disrupted and
in what way gas is expelled on a small scale.  This has important 
consequences for investigating the early evolution of star clusters.  A paper 
is in preparation on the effects of gas expulsion on the dynamics of the 
stellar content of a cluster and what star formation efficiency is 
required for a bound object to remain.

Using supercomputers $N=10^7$ or more is possible, giving the resolution 
to model fully the dynamical and chemical evolution of a dwarf 
galaxy.  In addition feedback can be placed into simulations of 
the formation of the first cosmological objects.

\section*{ACKNOWLEDGMENTS}

PAT is a PPARC Lecturer Fellow.  We would like to thank Hugh Couchman for 
useful discussions about this paper and acknowledge NATO CRG 970081 
which facilitated this interaction.  We would also like to thank Anne 
Green and Roger Hutchings for their help.  This project was 
completed using the computer facilities of the University of Sussex 
Astronomy Centre and University of Sussex BFG computer using the 
Hydra code (public release version available from the Hydra 
Consortium at http://phobos.astro.uwo.ca/hydra\_consort/).

In memory of Jenny.

\newpage

\begin{figure}
\centerline{\psfig{figure=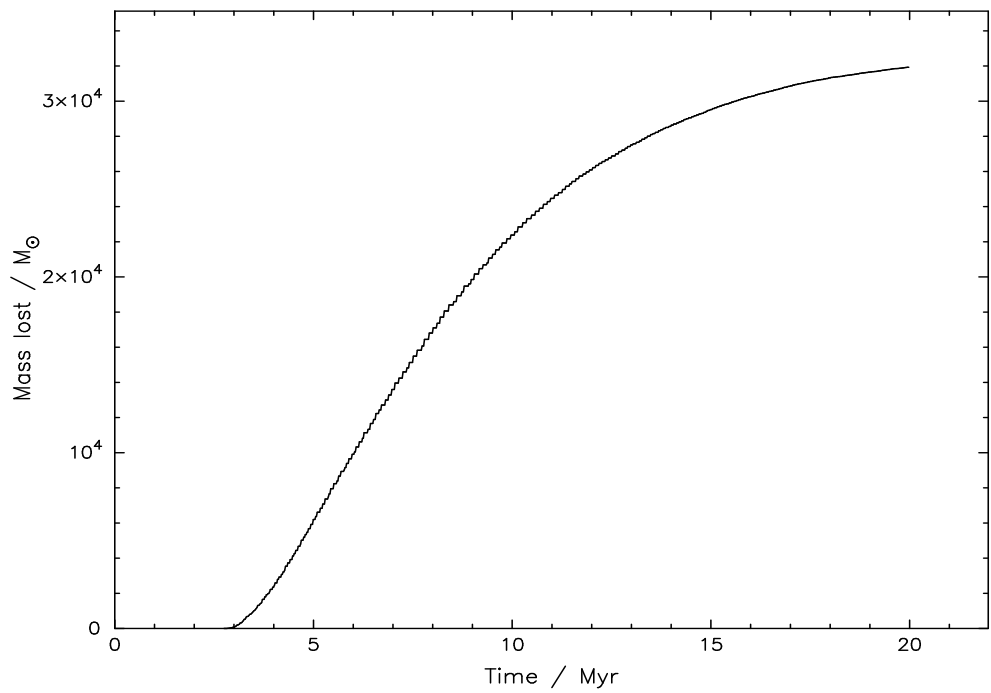,height=9.0cm,width=13.0cm,angle=270}}
\caption{The mass loss rate across the sides of the simulation cube
with time for a $M = 10^5 M_{\odot}$, $R=3.5$ pc, $r_{\rm max}=20$ pc Plummer cloud 
with a central supernovae.}
\label{fig:10to5mlost}
\end{figure}

\begin{figure}
\centerline{\psfig{figure=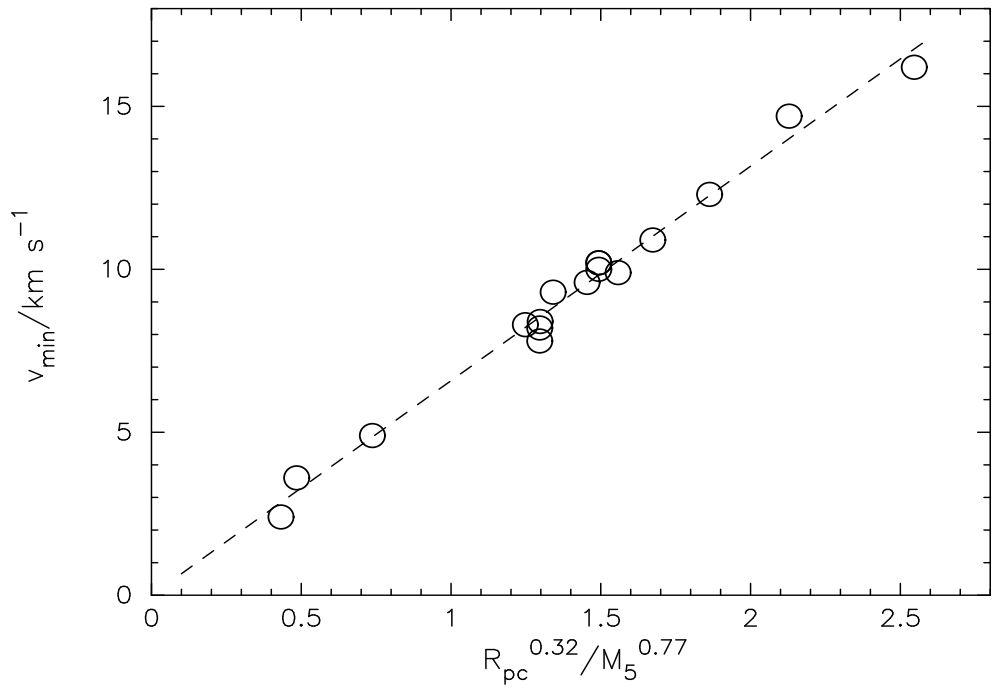,height=9.0cm,width=13.0cm,angle=270}}
\caption{The minimum shock velocity $v_{\rm min}$ in km s$^{-1}$ 
against the Plummer cloud characteristics $R_{\rm pc}^{0.32}/M_5^{0.77}$ 
from eqn.~\ref{eqn:vmin}.}
\label{fig:vminMR}
\end{figure}

\begin{figure}
\centerline{\psfig{figure=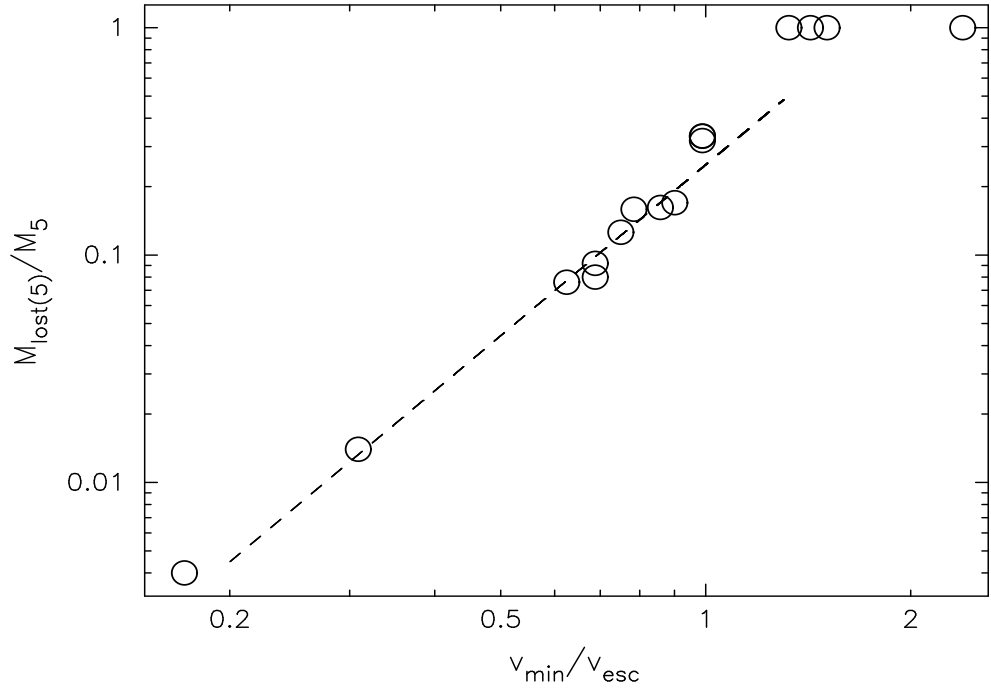,height=9.0cm,width=13.0cm,angle=270}}
\caption{The fractional mass lost against the ratio of the minimum 
velocity to escape velocity for the Plummer clouds listed in 
table~\ref{tab:models}.  A fit from eqn.~\ref{eqn:mlost} is added 
for clouds with $0.2 v_{\rm esc} < v_{\rm min} < v_{\rm esc}$ (clusters 
with partial mass loss - see text).}
\label{fig:mlost}
\end{figure}

\begin{figure}
\centerline{\psfig{figure=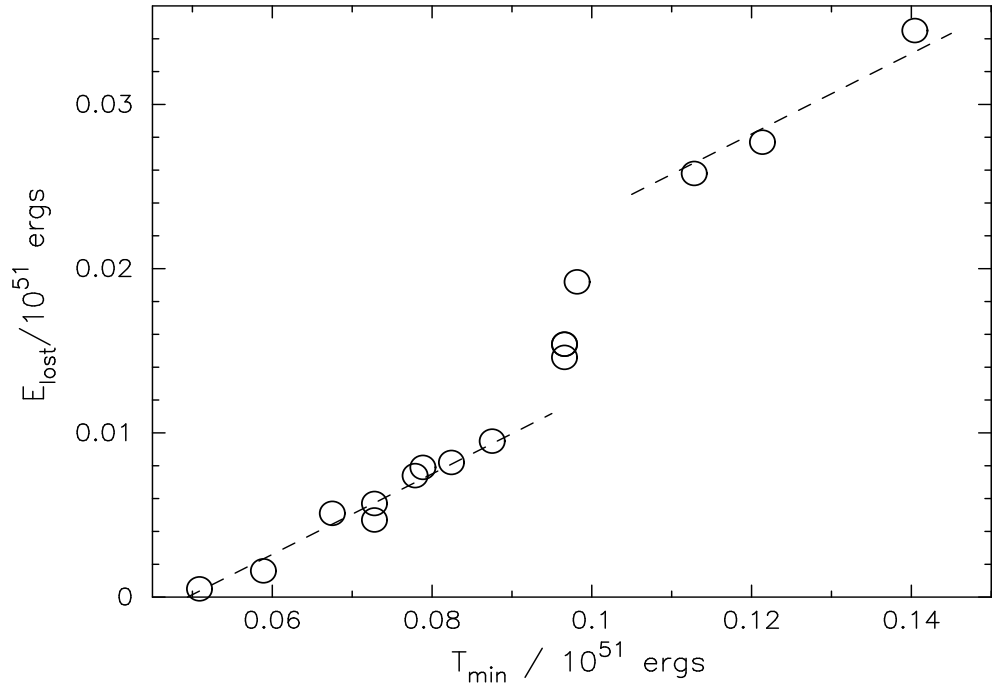,height=9.0cm,width=13.0cm,angle=270}}
\caption{The feedback energy $E_{\rm lost}$ against the minimum 
kinetic energy parameter $T_{\rm min} = 0.001 
M_5 (v_{\rm min}/$km s$^{-1})$.  Linear fits of the same slope are 
fitted from eqns.~\ref{eqn:elost1} \& ~\ref{eqn:elost2} are marked 
by dashed lines.}
\label{fig:elost}
\end{figure}

\begin{figure}
\centerline{\psfig{figure=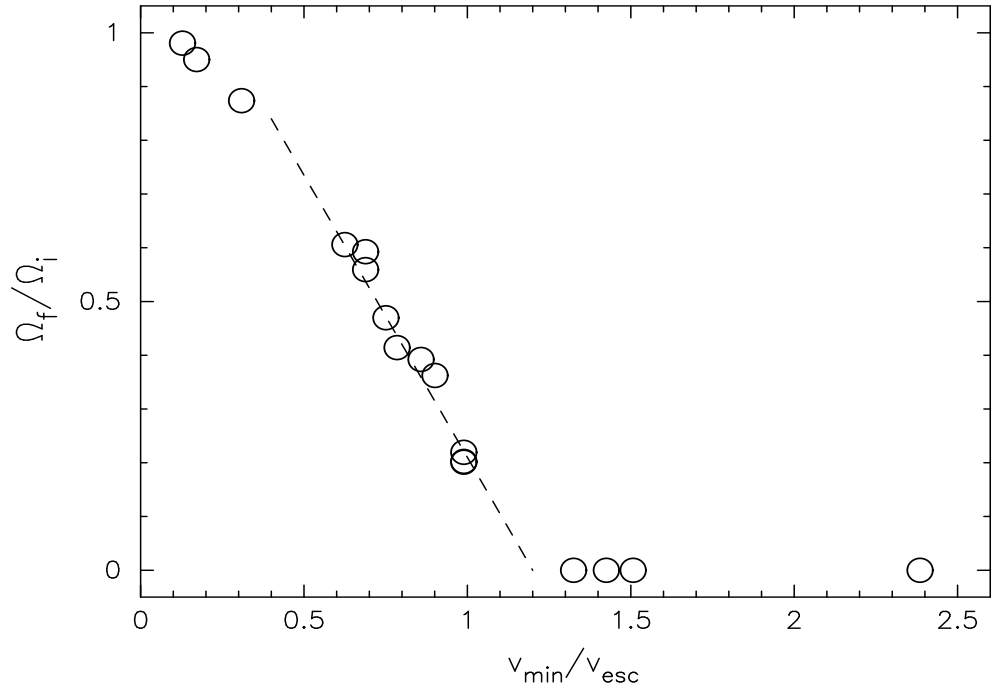,height=9.0cm,width=13.0cm,angle=270}}
\caption{The final-to-initial potential energy ratio against the ratio of 
the minimum velocity to escape velocity for the Plummer clouds listed in 
table~\ref{tab:models}.  The linear relationship marked by the 
dashed line is from eqn.~\ref{eqn:potlost}.}
\label{fig:potlost}
\end{figure}

\begin{figure}
\centerline{\psfig{figure=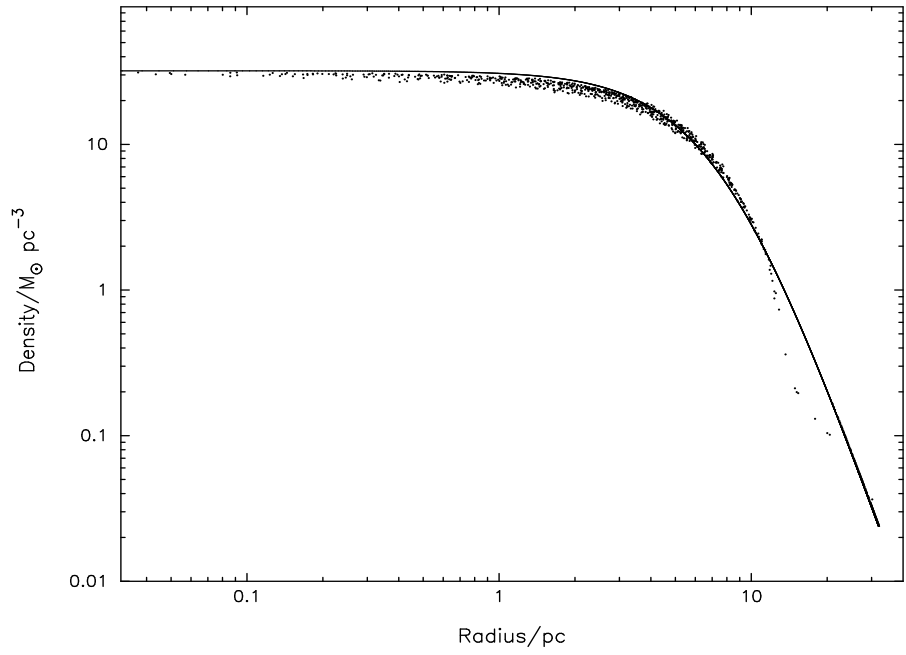,height=9.0cm,width=13.0cm,angle=270}}
\caption{The density distribution after 20 Myr for the $10^5 M_{\odot}$ Plummer 
cloud from fig.~\ref{fig:10to5mlost}.  The solid line is a Plummer model 
with $M=6.36 \times 10^4 M_{\odot}$ and $R=7.8$ pc}
\label{fig:final}
\end{figure}

\newpage

\begin{deluxetable}{ccccccc} 
\footnotesize
\tablecaption{Parameters of the ten $n_0 = 0.5$ cm$^{-3}$, $\zeta_m = 1$
convergence test runs.  The table gives the run identification number, box 
size (in pc), softening (in pc), number of particles $N$, the end time of 
the simulation (in Myr), the mass of an individual particle (in $M_{\odot}$) and 
the symbol used to represent each run in figs.~\ref{fig:alle},\ref{fig:energyshells},\ref{fig:revol} and 
\ref{fig:snrcomp}.}
\tablehead{
\colhead{RUN} & \colhead{box} & \colhead{softening} & \colhead{$N$} & \colhead{End time} & \colhead{Particle mass} & \colhead{Symbol} \nl 
 & \colhead{pc} & \colhead{pc} & & \colhead{Myr} & \colhead{$M_{\odot}$} &  
}
\startdata
1001 & 300 & 27 &          $32^3$ & 3 & 11.1 & dot \nl
1002 & 300 & 23 & $2 \times 32^3$ & 3 & 5.53 & cross \nl
1003 & 300 & 18 & $4 \times 32^3$ & 3 & 2.77 & star \nl
1004 & 300 &  5 &          $64^3$ & 2 & 1.38 & circle \nl
1005 & 300 &  5 & $2 \times 32^3$ & 3 & 5.53 & 'x' \nl
1017 & 200 & 11 & $4 \times 32^3$ & 2 & 0.82 & square \nl
1018 & 400 & 20 & $4 \times 32^3$ & 5 & 6.56 & triangle \nl
1019 & 500 & 28 & $4 \times 32^3$ & 6 & 12.8 & diamond \nl
1020 & 500 & 22 &          $64^3$ & 6 & 6.40 & open star \nl
1021 & 200 &  9 &          $64^3$ & 2 & 0.41 & open cross \nl
\enddata
\label{tab:runs}
\end{deluxetable}

\begin{deluxetable}{ccccc} 
\footnotesize
\tablecaption{The density of the surrounding ISM, the initial radius of 
the PDS and the approximate time and radius at which the shell velocity 
falls below the ambient medium's sound speed. The evolution of the shell 
radiuswith time for the different ambient densities used here is  
plotted in fig.~\ref{fig:diffdens}.}
\tablehead{
\colhead{RUN} & \colhead{$n_0$} & \colhead{$r_{\rm PDS}$} & \colhead{$t_{\rm sound}$} & \colhead{$r_{\rm sound}$} \nl 
& \colhead{cm$^{-3}$} & \colhead{pc} & \colhead{Myr} & \colhead{pc} 
}
\startdata
1006 & 0.01 & 101 & 7.2  & 360 \nl
1007 & 0.1  & 38  & 3.0  & 150 \nl
1008 & 0.2  & 28  & 2.2  & 120 \nl
1003 & 0.5  & 18  & 1.6  & 84  \nl
1009 & 1    & 14  & 1.3  & 66  \nl
1010 & 5    & 7.0 & 0.70 & 37  \nl
1011 & 10   & 5.2 & 0.45 & 26  \nl
1012 & 100  & 1.9 & 0.21 & 11  \nl
\enddata
\label{tab:sound}
\end{deluxetable}

\begin{deluxetable}{cccccc} 
\footnotesize
\tablecaption{The cloud models and feedback parameters of the runs in 
Section 3. The Plummer parameters mass $M_5$ and radius $R_{\rm pc}$  
for each cloud are followed by the minumum velocity $v_{\rm min}$ of the 
shell in that cloud, the fractional mass loss $M_{\rm lost}/M$, the 
energy feedback $E_{\rm lost}$ and the loss of potential energy 
$\Omega_{\rm lost}$ from that cloud.}
\tablehead{
 \colhead{$M$} & \colhead{$R$} & \colhead{$v_{\rm min}$} & \colhead{$M_{\rm lost}/M$} & \colhead{$E_{\rm lost}$} & \colhead{$\Omega_{\rm lost}$} \nl 
 \colhead{$10^5 M_{\odot}$} & \colhead{pc} & km s$^{-1}$ & & \colhead{$10^{48}$ ergs} & \colhead{$10^{48}$ ergs} 
}
\startdata
 0.5  &  2  & 14.7 & 1.00 & 19  & 32  \nl
 0.5  & 3.5 & 16.2 & 1.00 & 35  & 18  \nl
 0.75 &  2  &  9.9 & 0.17 & 7.9 & 45  \nl
 0.75 & 3.5 & 12.3 & 1.00 & 26  & 40  \nl
 0.9  & 2.5 &  9.6 & 0.16 & 8.2 & 50  \nl
 1    &  2  &  8.3 & 0.076 & 5.1 & 50  \nl
 1    & 2.25&  7.8 & 0.080 & 4.7 & 45  \nl
 1    & 2.25&  8.2 & 0.092 & 5.7 & 49  \nl
 1    & 2.5 &  9.3 & 0.13 & 7.4 & 53  \nl
 1    & 3.5 & 10.2 & 0.33 & 15 & 57  \nl
 1    & 3.5 & 10.0 & 0.32 & 15 & 56  \nl
 1    & 3.5 & 10.2 & 0.33 & 15 & 57  \nl
 1    &  5  & 10.9 & 1.00 & 28 & 49  \nl
 1.2  & 3.5 &  8.4 & 0.16 & 9.5 & 60  \nl
 2.5  & 3.5 &  4.9 & 0.014 & 1.6 & 56  \nl
 5    & 3.5 &  2.4 & 0    & 0    & 34  \nl
 5    & 5   &  3.6 & 0.004 & 0.5 & 62  \nl
\enddata
\label{tab:models}
\end{deluxetable}

\end{document}